\documentstyle[floats,aps,twocolumn]{revtex}

\begin{document}

\twocolumn[\hsize\textwidth\columnwidth\hsize\csname@twocolumnfalse\endcsname
\title{Gardiner's phonon for Bose-Einstein Condensation: 
A physical realization of the $q$-deformed Boson }
\author{Chang-Pu Sun\cite{email}\cite{www}, Sixia Yu}
\address{Institute of theoretical Physics, Academia Sinica, 
Beijing 100080, P.R. China}
\author{Yi-Bao Gao}
\address{Department of Physics, Beijing Polytechnic University, 
Beijing 100022, P.R. China}

\maketitle

\begin{abstract}
The Gardiner's phonon presented for a particle-number conserving 
approximation method to describe the dynamics of  Bose-Einstein 
Condesation (BEC) (C.W. Gardiner, Phys. Rev. A 56, 1414 (1997)) 
is shown to be a physical realization of the $q$-deformed boson,
which was abstractly developed in quantum group theory. On this 
observation, the coherent output of BEC atoms driven by a radio
frequency (r.f) field is analyzed in the viewpoint of a $q$-deformed 
Fock space. It is illustrated that the $q$-deformation of bosonic 
commutation relation corresponds to the non-ideal BEC with the 
finite particle number $N$ of condensated atoms. Up to order $1/N$, 
the coherent output state of the untrapped atoms {\it minimizes} 
the uncertainty relation like a coherent state does in the ideal 
case of BEC that $N$  approaches infinity or  $q=1-2/N$ approaches 
one.
 
\end{abstract}
]

\section{Introduction}

Since the introduction of the concept of $q$-deformed boson \cite{q1} by
different authors independently in the development of the quantum group
theory \cite{q2,q3,q4}, there have been many efforts of finding its physical
realizations \cite{q5,green}, e.g., fittings the deformed spectra of
rotation and oscillation for molecules and nuclei \cite{bonat,chang}. In our
opinion, those investigations can be regarded as merely phenomenological
because a $q$-deformed structure is postulated in advance without giving it
a microscopic mechanism. In this paper it will be shown that a physical and
natural realization of the $q$-deformed boson is provided by the phonon
operators, which was presented recently by Gardiner \cite{Gardiner} for the
description of Bose-Einstein condensation (BEC).

As well known, to deal with the dynamics of Bose-Einstein condensated gas
the Bogoliubov approximation in quantum many-body theory is an efficient
approach, in which the creation and annihilation operators for condensated
atoms are substituted by $c$-numbers. One shortcoming of this method is that
the total atomic particle-number may not be conserved after the
approximation. Or a symmetry may be broken. To remedy this default, Gardiner
suggested a modified Bogoliubov approximation by introducing phonon
operators which conserve the total atomic particle number $N$ and obey the
bosonic commutation relations in case of $N{\to \infty }$. In this sense
this phonon operator approach gives an elegant infinite atomic particle
number approximation theory for BEC taking into account the conservation of
the total atomic number.

What will be investigated here is the case that the total atomic particle
number $N$ is very large but not infinite. That is, we shall consider the
effects of order $o(1/N)$. And we shall focus on an algebraic method of
treating the effects of finite condensated particle number in atomic BEC. As
it turns out, the commutation relations for the Gardiner's phonon operators
will no longer obey the commutation relation of the Heisenberg-Weyl algebra
but the $q$-deformed bosonic commutation relation 
\begin{equation}
\left[ b,b^{\dagger }\right] _q\equiv bb^{\dagger }-qb^{\dagger }b=1,
\label{1}
\end{equation}
where the deformation constant $q$ depends on the total atomic particle
number. By using this $q$-deformed boson algebra we will discuss the effects
of the finiteness of the particle number on the output of MIT experiment of
BEC.

This paper is organized as follows. In sec.II after the reformulation of the
Gardiner's modified Bogoliubov method for the two-level atoms, the $q$%
-deformed commutation relation for the Gardiner's phonon operator is derived
in case of large but finite particle number. The excitation of the
Gardiner's phonons is also considered for small $N$. In sec.III, the
Hamiltonian describing the coherent output in MIT experiment for atomic
laser \cite{mew} is discussed in terms of the $q$-deformed algebra. When $%
q=1 $ or $N\to \infty $ it describes the coherent output of BEC from a
vacuum to a coherent states. For general $q$, or finite but large
condensated atomic particle number, it rusults in a nonlinear dynamic
Heisenberg equation for the $q$-deformed bosonic operators. Its solution at
the first order of $1/N$ is analyzed. In sec.IV dressed phonon, a
generalization of the Gardiner's phonon in case of quantized radiation
field, is disscussed and it satisfies also a $q$-deformed bosonic algebra
for a large phonton number and atomic particle number. Finally there are
some concluding remarks in the last section.

\section{$q$-deformed bosonic algebra for the Gardiner's phonon}

Gardiner's starting point to introduce the phonon operators is to consider a
system of the weakly interacting Bose gas. Without losing generality, we
consider for the moment a classical radiation field interacting with
two-level atoms with $b_e^{\dagger }$ and $b_e$ denoting the creation and
annihilation operators for the atoms in the untrapped state and $%
b_g^{\dagger }$ and $b_g$ for the creation and annihilation operators of the
atoms in the trapped ground state. The simplified Hamiltonian for the MIT
experiment reads 
\begin{equation}
H=\hbar \omega _eb_e^{\dagger }b_e+\hbar g(t)b_e^{\dagger }b_g+{\rm h.c.},
\label{h1}
\end{equation}
where $g(t)$ is a time-dependent coupling coefficient for the classical
sweeping r.f. field coupled to those two states with level difference $\hbar
\omega _e$. Note that the total atomic particle number ${N}%
=b_e^{\dagger}b_e+b_g^{\dagger }b_g$ is conserved. For convenience we define 
$\eta =1/{N}$ for large particle number.

In the thermodynamical limit ${N}\to \infty $, the Bogoliubov approximation
is usually applied, in which the ladder operators $b_g^{\dagger },b_g$ of
the ground state are replaced by a $c$-number $\sqrt{N_c}$, where $N_c$ is
the average number of the ininital condensated atoms. As a result
Hamiltonian Eq.(\ref{h1}) becomes a forced harmonic oscillator with the
external force term $\hbar \sqrt{N_c}g(t)b_e^{\dagger }+{\rm h.c}$.
Obviously, it will drive the atom in the untrapped excited state from a
vacuum state to a coherent state and thus lead to a coherent output of BEC
atoms in the propagating mode . In this case the condensated atoms are
initially described by a coherent state\cite{mew}.

However, this apporoximation destroyes a symmetry of the Hamiltonian Eq.(\ref
{h1}), i.e., the conservation of the total particle number is violated
because of $[{N},H_b]\neq 0$. According to Gardiner, the phonon operators
are defined as: 
\begin{equation}
b=\frac 1{\sqrt{{N}}}b_g^{\dagger }b_e,\qquad b^{\dagger }=\frac 1{\sqrt{{N}}%
}b_gb_e^{\dagger }.
\end{equation}
These operators act invariantly on the subspace $V^N$ spanned by bases $%
|N;n\rangle \equiv |N-n,n\rangle $ $(n=0,1,\ldots ,N)$, where Fock sates 
\[
|m,n\rangle =\frac 1{\sqrt{m!n!}}b_e^{\dagger m}b_g^{\dagger n}|0\rangle 
\]
with $m,n=0,1,2,\ldots $ spann the Fock space $H{}_{2b}$ of a two mode boson.

A straightforward calculation leads to the following commutation relation
between the phonon operaotr and it Hermitian conjugate: 
\begin{equation}
\left[ b,b^{\dagger }\right] =1-\frac 2{{N}}b_e^{\dagger }b_e=f(b^{\dagger
}b;\eta ),
\end{equation}
with $f(x;\eta )=\sqrt{1+2(1-2x)\eta +\eta ^2}-\eta$. Keeping only the
lowest order of $\eta $ for a very large total atomic particle number, the
commutator above becomes 
\begin{mathletters}
\begin{equation}
\left[ b,b^{\dagger }\right] =1-2\eta b^{\dagger }b
\end{equation}
or 
\begin{equation}
\left[b,b^{\dagger }\right] _q=1,  \label{bcr}
\end{equation}
with $q=1-2\eta$. This is exactly a typical $q$-deformed commutation
relation. As $N\to\infty $ or $q\to 1$, the usual commutation relation of
Heisenberg-Weyl algebra is regained.

Since the Gardiner's phonon excitation for BEC can be mathematically
understood in terms of $q$-deformed boson system, the algebraic method
previously developed in quantum group theory can be of use for the
construction of a $q$-deformed Fock space. The existence of a vacuum state $%
|0\rangle $ such that $b|0\rangle =0$ is guaranteed by the fundamental
structure of the $q$-deformed boson algebra \cite{sun1}. Starting with the
vacuum, from relation 
\end{mathletters}
\begin{equation}
bb^{\dagger ^n}=q^nb^{\dagger ^n}b+\langle n\rangle b^{\dagger n-1},
\end{equation}
where the $q$-deformed $c$-number $\langle n\rangle =n-n(n-1)\eta )$ is up
to the first order of $\eta $ and approaches to $n$ when $N\to \infty $, the 
$q$-deformed Fock space for the Gardiner's phonon excitation of BEC is
spanned by bases 
\begin{equation}  \label{base}
|n\rangle =\frac 1{\sqrt{\langle n\rangle !}}b^{\dagger n}|0\rangle
\end{equation}
with $\langle n\rangle!=\langle n\rangle \langle n-1\rangle \cdots \langle
1\rangle $ $(n=0,1,\ldots,N)$. The action of operators $b$ and $b^{\dagger }$
are then represented on the deformed Fock space as: 
\begin{equation}
b^{\dagger }|n\rangle =\sqrt{\langle n+1\rangle }|n+1\rangle ,\quad
b|n\rangle =\sqrt{\langle n\rangle }|n-1\rangle
\end{equation}
the actions of the creation and annihilation operators $b$ and $b^{\dagger }$
coincide with their direct actions on the subspace $V^N$ of the Fock space
of a two-mode boson on which the total particle number is conserved.

Actually, the $q$-deformed vacuum state is physically a two-mode boson state 
$|N;0\rangle $ and it is trivially annihilated by the phonon operator $b$.
So the vacuum state can be regarded as a case that all bosons condensed in
the ground state and tends to form a BEC. By mapping back to two-mode boson
space, the $q$-deformed Fock state $|n\rangle \sim |N;n\rangle$ implies a
phonon excitation that $n$ bosons are created in the propagating mode by
destroying n bosons in trapped mode. This kind of fundamental process of
phonon excitation essentially guaranteed the particle number conservation.

In the above discussion about the Gardiner's phonon excitation, we have
linearized commutator $h\equiv f(b^\dagger b;\eta)$ so that a $q$-deformed
commutation rule was obtained. Essentially this linearization establishes a
physical realization of the $q$-deformed algebra. However, if the total
particle number $N$ is not large enough, then $h$ can not be approximated by
a linear function. From the commutation relations between $h$ and $%
b,b^{\dagger }$%
\begin{equation}
\left[ h,b^{\dagger }\right]=-\frac 2Nb^{\dagger },\qquad [h,b]=\frac 2Nb,
\end{equation}
we see that the algebra of Gardiner's operators is a rescaling of algebra $%
su(2)$ with factor $N$. When $N\to\infty$, it approaches the Heisenberg-Weyl
algebra because $h$ approaches identity. Its representation space is a
finite dimensional space is found to be spanned by bases given in Eq.(\ref
{base}) with $\langle n\rangle=n-n(n-1)\eta$ being exact. It is interesting
to notice that $\langle n\rangle$ coincide with the $q$-deformed number in
the first order of $\eta$. The action of those creation and annihlation
operators have the same form. On this finite dimensional space we have $%
b|0\rangle =0$ and $h|0\rangle=|0\rangle $ together with $b^{\dagger
}|N\rangle =0$. It is worthy to point out that the last equality can not be
composed to the usual Heisenberg-Weyl algebra because such a constrain will
make it is impossible for the creation operator to be the Hermitian
conjugate to the annihilation operator. Physically the usual boson algebra
and its $q$-deformation with q being not a root of unity describe systems
without particle number conservation such as the black body radiation, but
this finite dimensional representation provide us with a suitable algebraic
framework to treat the dynamic process where the particle number is
conserved.

\section{The coherent output of BEC as a $q$-deformed bosonic excitation}

In this section we shall apply the concept of the $q$-deformed boson algebra
for the Gardiner's phonon excitation to the theoretical analysis of the MIT
experiment for the coherent output of the BEC atoms from a trap.

By denoting $N_e=b_e^{\dagger }b_e$, in terms of the deformed boson
operators the Hamiltonian can be rewritten as: 
\begin{equation}
H_{q}=\hbar \omega _eN_e+\hbar \sqrt{{\bf N}}\left(g(t)b^\dagger+{\rm h.c.}%
\right).
\end{equation}
For an ideal BEC with infinite atoms condensated in the trapped state, the $%
q $-deformed bosonic commutation relation becomes the usual commutation
relation $[b,b^{\dagger }]=1$ of HW algebra. Together with commutator $%
[N_e,b]=-b$, a linear Heisenberg equation 
\begin{equation}
\dot{b}=-i\omega _eb-i\mu(t)
\end{equation}
is acquired where $\mu(t)=g(t)\sqrt{{N}}$ is finite as $N$ tends to
infinity. Its solution can be easily found to be 
\begin{equation}
b(t)=b(0)e^{-i\omega _et}+\beta (t)
\end{equation}
with a non-vanishing expectation value in the vacuum state: 
\begin{equation}
\beta (t)=\langle 0|b(t)|0\rangle= -i\int_0^te^{-i\omega _e(t-t^{\prime
})}\mu(t^{\prime})dt^{\prime }.
\end{equation}

Therefore when the BEC is initially at a Fock state, one obtains a coherent
state output of the untrapped atoms. This is just a direct manifestation of
the quantum coherence through a factorization $\langle
b^\dagger(t)b(t)\rangle=\langle b^\dagger(t)\rangle\langle b(t)\rangle$. In
fact for infinite $N$, the average phonon number $\langle b^{\dagger
}(t)b(t)\rangle $ approach the average atomic number $b_e^{\dagger }b_e$ of
outputting in the untrapped mode.

Further, to consider the quantum coherence of the outputting atoms
represented by phonon operators $b(t)$, we evaluate the variances of the
Hermitian quadrature phase operators \cite{sun3} 
\begin{equation}
X_1(t)=\frac{b(t)+b^{\dagger }(t)}{\sqrt{2}},\quad X_2(t)=\frac{%
b(t)-b^{\dagger }(t)}{i\sqrt{2}}.
\end{equation}
One can easily find that 
\begin{equation}
\Delta ^2X_{1,2}(t)=\stackrel{}{\langle 0|X_{1,2}^2|0\rangle }-\langle
0|\Delta X_{1,2}^{}|0\rangle ^2=\frac 12
\end{equation}
are time-independent constants that minimize the uncertainty product $%
\Delta^2X_1(t)\Delta^2X_2(t)=1/4$. It reflects that the outputted atoms are
just in a coherent state. This judgements for the case with infinite $N$ can
be similarly used to analyses the influence of $q$-deformation with finite
atomic number on the dynamics of outputting BEC.

In practice condensated atomic number $N$ cannot be infinite and therefore
the Bogoliubov approximation can not work well and so an approximation
method with higher accuracy that for the generic Bogoliubov approximation or
its Gardiners modification. And the $q$-deformed boson algebra disscussed in
the previous section provides us with a algebric structure of treating this
correction of finite atomic particle number.

Up to the approximation of order $\eta$, as discussed previously, the
dynamics of Gardiner's phonon excitation for coherent output of BEC is
determined by the $q$-deformed boson commutation relation. The corresponding
Heisenberg equation 
\begin{equation}
\dot{b}(t)=-i\omega _eb(t)-i\mu (t)+2i\eta \mu (t)b^{\dagger }(t)b(t)
\label{he}
\end{equation}
is then a nonlinear one. It is should be remarked here that that the
effective coupling $\mu (t)=g(t)\sqrt{N}$ is finite in the thermodynamical
limit of an infinite number of atoms in an infinite volume but with the
density fixed, i.e., $n=\lim_{N,V\to \infty}\left( N/V\right)$ is finite.
This is because the coupling constant $g(t)$ between the r.f. pulse and
atoms must be proportional to the inverse of the square root of the the
effective mode volume $V$. So we have $\mu (t)\propto \sqrt{N/V}$, the
square root of the density of the condensated atoms in the thermodynamical
limit. For this reason the nonlinear terms in Eq.(\ref{he}) can be handled
as a perturbation in comparison with other terms. Otherwise, if $\mu (t)$
were not fixed as $N$ approaches infinity, the non-liner term should not be
regarded as a perturbation due to its square amplification of perturbation.
The thermodynamical limit is our key point to deal with the dynamics of
q-deformation for the coherent output of the BEC atoms with a perturbation
method.

Becasue of the nature of the deformed bonon, in general, it is only
neccissary to solve this nonlinear Heisenberg equation Eq.(\ref{he})
according to the first order of $\eta$ by perturbation. In fact if the
annihilation operator at time $t$ is expanded as 
\begin{equation}
b(t)= b_0e^{-i\omega_et}+\beta (t)+\eta b_1(t)
\end{equation}
with $b_0=b(0)$, one obtains immediately 
\begin{mathletters}
\begin{eqnarray}
&b_1(t)=\displaystyle\xi(t)b_0+\beta^2(t)e^{i\omega_et}b^{\dagger }_0
-2\beta(t)b^{\dagger }_0b_0+\alpha (t),& \\
&\alpha (t)= \displaystyle 2i\int_0^te^{-i\omega_e(t-t^{\prime })}\mu
(t^{\prime })|\beta (t^{\prime })|^2dt^{\prime },& \\
&\xi (t)=\displaystyle 2i\int_0^te^{-i\omega_e t}\mu (t^{\prime })\beta
^*(t^{\prime })dt^{\prime }.&
\end{eqnarray}

As a result the average outputted atomic number $\langle N_e(t)\rangle $ in
the untrapped excited state with finite condenstated BEC particle number
correction is given by 
\end{mathletters}
\begin{equation}
\langle N_e(t)\rangle =|\beta (t)|^2+\eta \left( \left| \beta (t)\right| ^4+2%
{\rm Re}\left( \beta (t)\alpha (t)^{*}\right) \right) .
\end{equation}
The quantum fluctuation of the quadrature phase operators $X_1(t)$ and $%
X_2(t)$ in the final state can also be explicitly determined as: 
\begin{mathletters}
\begin{eqnarray}
\Delta ^2X_1 &=&\displaystyle\frac 12-\eta \left( \left| \beta (t)\right|
^2+\beta ^2(t)+\beta ^{*2}(t)\right) ,  \label{fl1} \\
\Delta ^2X_2 &=&\displaystyle\frac 12-\eta \left( \left| \beta (t)\right|
^2-\beta ^2(t)-\beta ^{*2}(t)\right) .
\end{eqnarray}
Keeping the first order of $\eta $, one obains immediately the follwing
relation 
\end{mathletters}
\begin{equation}
\Delta ^2X_1\Delta ^2X_2=\frac 14-\eta \left| \beta (t)\right| ^2.
\end{equation}
The term of $\eta$ reflects the devotion from the thermodynamical limit as
the effect of the $q$-deformation with finite condensated particle number.

Since the commutation relation of quadrature phase operators $X_1(t)$ and $%
X_2(t)$ is deformed into 
\begin{equation}
\lbrack X_1(t),X_2(t)]=i\left( 1-2\eta b^{\dagger }(t)b(t)\right),
\end{equation}
the corresponding Heisenberg's uncertainty relationship of those quadratures
in the output taking account of finite atomic number should be: 
\begin{equation}
\Delta ^2X_1\Delta ^2X_2\geq \frac 14-\eta \langle 0|b^{\dagger
}(t)b(t)|0\rangle.
\end{equation}
Considering $\langle 0|b^{\dagger }(t)b(t)|0\rangle =\left| \beta(t)\right|
^2+o(\eta)$ the coherent output state of the untrapped atoms {\it minimizes}
the uncertainty relation. Because, as we known, the squeezed state and
coherent state can be defined to be states minimize the uncertainty relation
between the position and momentum, the output state of the atomic lase can
be likened to a kind of $q$-deformed coherent state or even a squeezed state
because fluctuations Eq.(26a) and Eq.(26b) are inequal.

\section{Dressed phonon operator}

When the r.f field is a quantized one which is described by a single-mode
boson whose annihilation operator is denoted as $a$, we consider the
following Hamiltonian: 
\begin{equation}
H_d=\omega _eN_e+\omega _gN_g+\omega _0N_0+g(a^{\dagger }b_g^{\dagger
}b_e+ab_gb_e^{\dagger }),  \label{hd}
\end{equation}
where $N_0=a^{\dagger }a$ is the photon number operator and $g$ is a real
coupling constant. Obviously there is a conservative quantity $%
\Delta=N_0+N_e $ in addition to ${N}$. We now define the dressed phonon
operators as 
\begin{equation}
B=\frac 1{\sqrt{{N\Delta }}}a^{\dagger }b_g^{\dagger }b_e,\qquad B^{\dagger
}=\frac 1{\sqrt{{N\Delta }}}ab_gb_e^{\dagger }.
\end{equation}
For convenience we denote $\eta _0=1/\Delta $ for large $\Delta $. The
dressed phonon operators satisfy the following commutation relation: 
\begin{equation}
\left[ B,B^{\dagger }\right] =1-2(\eta _0+\eta )N_e+\eta _0\eta (3N_e^2-N_e).
\end{equation}
In terms of those dressed phonon operators the Hamiltonian Eq.(\ref{hd}) can
be rewritten as 
\begin{equation}
H_d=\omega _g{N}+\omega _0\Delta +\omega _\delta N_e+\mu _d(B+B^{\dagger }).
\end{equation}
where $\mu _d=g\sqrt{{N}\Delta }$ and $\omega _\delta =\omega _e-\omega
_g-\omega _0$.

As $N$ (or $\Delta$) tends to infinity the dressed phonon operators $%
B,B^{\dagger }$ and $N_e$ will give a representation of algebra $su(2)$. As
both $N$ and $\Delta$ tends infinity one obtains Heisenberg-Weyl algebra $%
[B,B^{\dagger }]=1$. When we keep the first order in $o(\eta )$ and $o(\eta
_0)$, and considering $B^{\dagger }B=N_e+o(\eta +\eta _0)$, we obtain a $q$%
-deformed boson algebra 
\begin{equation}
\left[ B,B^{\dagger }\right] _{q_d}=1  \label{hed}
\end{equation}
with the deforming constant given by $q_d=1-2(\eta +\eta _0)$.

Within this kind of approximation, the Heisenberg equation of motion
satisfied by the dressed phonon is given by: 
\begin{equation}
\dot B=-i\omega_\delta B-i\mu_d+2i(\eta_0+\eta)\mu_dB^\dagger B.
\end{equation}
Here the coupling constant $g$ is proportional to the inverse of the square
root of volume $V$ which makes $\mu_d$ finite as ${N}$ and $\Delta$
approaches infinity. Because the evolution equation Eq.(\ref{hed}) is the
same with Heisenberg equation Eq.(\ref{he}), all the results discussed in
the previous section should also be valid in case of quantized r.f. field.

\section{Remarks}

To conclude this paper, we treat the same problem in the Schr\"odinger
picture. Without the effect of $q$-deformation caused by finite $N$, the
dynamic process can be described as a factorized evolution of the wave
function\cite{sun3}. Suppose the initial state is $|\phi (0)\rangle =|\alpha
=\sqrt{N}\rangle _g\otimes|0\rangle _e$ i.e., all atoms occupied the trapped
groud state with a coherent state while there is no atom in the propagating
mode. Driven by a r.f. field $g(t)=g\exp(i\omega _ft)$ ($g\propto \sqrt{1/V}$%
) with a fixed frequency $\omega _f$, the system will have exactly a
factorized evolution wave function of the form $|\phi (t)\rangle =|\alpha
\rangle _g\otimes |\beta (t)\rangle _e$ even for a finite $N$. Such a
factorization structure of wave function first pointed out by Mewes et al is
very crucial to realize their experiment, and its dynamical origins mainly
depend on the linearity of the original coupling system of $b_e$ and $b_g$.
The coherent state of the second component $|\beta (t)\rangle _e$ of $|\phi
(t)\rangle $ means a coherent output of atoms in propagating mode in the MIT
experiment. The influence of finite $N$ on Schroedinger wave function is
manifested by the outputting amplitude $|\beta(t)|\propto\sqrt{N/V}|\sin
(\Omega t)|$ where $\Omega =\sqrt{(\omega-\omega _f)^2/4+g^2}$. Because the
factor $\sqrt{N/V}$ is finite and $\Omega$ tends to $|\omega -\omega _f|$ in
the thermodynamics limit, the results from the Bogoliubov approximation
mentioned in Sec.II is just recovered by this exact solution.

There is another nonlinear effect different from that resulting from the $q$%
-deformation with finite $N$. In presence of interatomic interaction in the
trapped or untrapped states, which roughly is of the form $%
b_i^{\dag}b_j^{\dag }b_kb_l$, the corresponding Heisenberg equations are no
longer linear and thus the factorization structure will be broken down. It
is noticed that such a non-linearity is essentially different from that
resulting from the q-deformation with finite $N$ in this paper. The later
originates from the subsystem isolation of a large system.

\vskip 0.3cm

\section*{Acknowledgment}

This work is partly supported by the NFS of China.

\newpage

\end{document}